\documentclass[12pt]{article}

\usepackage[cp1251]{inputenc}
\usepackage[russian]{babel}

\usepackage{graphicx,comment}
\usepackage{amsmath,amssymb,amsfonts,amsthm,comment}
\usepackage{array}

\begin{document}

\noindent {Preliminary version. Problems of Information Transmission, 2021}.

\vskip 0.4cm

\begin{center}
{\large ON MINIMAX DETECTION OF GAUSSIAN STOCHASTIC SEQUENCES AND GAUSSIAN
STATIONARY SIGNALS \footnote[1]{This work was supported  by the Russian Foundation
for Basic Research under Grant 19-01-00364.}} 
\end{center}

\begin{center} {\large M. V. Burnashev \footnote[2]
{Burnashev M. V. is with Institute for Information Transmission
Problems, Russian Academy of Sciences, Moscow; email: burn@iitp.ru}}
 \end{center}


\vskip 0.4cm

{\bf Abstract -- }
Minimax detection of Gaussian stochastic sequences (signals) with unknown
covariance matrices is studied. For a fixed false alarm probability 
(1-st kind error probability), the performance of the minimax detection is 
being characterized by the best exponential decay rate of the miss probability 
(2-nd kind error probability) as the length of the observation interval tends 
to infinity. Our goal is to find the largest set of covariance matrices such 
that the minimax robust testing of this set (composite hypothesis) can be 
replaced with testing of only one specific covariance matrix (simple hypothesis)
without any loss in detection characteristics. In this paper,
we completely describe this maximal set of covariance matrices.
Some  corollaries address minimax detection of the Gaussian
stochastic signals embedded in the White Gaussian noise and
detection of the Gaussian stationary signals.


\section{Introduction, Definitions and Main Results}
We consider the problem of the minimax testing of the simple
hypothesis $\mathcal H_{0}$ against a composite alternative
$\mathcal H_{1}$, based on observations
${\mathbf y}_{n}^{T} = {\mathbf y}_{n}' =
(y_{1},\ldots,y_{n}) \in {\mathbf R}^{n}$:
\begin{equation}\label{mod1}
\begin{split}
&\mathcal H_{0}: {\mathbf y}_{n} = \boldsymbol{\xi}_{n}, \qquad
\boldsymbol{\xi}_{n} \sim
{\mathcal N}(\boldsymbol{0},\mathbf{I}_{n}), \\
&\mathcal H_{1}: {\mathbf y}_{n} = \boldsymbol{\eta}_{n}, \qquad
\boldsymbol{\eta}_{n} \sim {\mathcal N}(\boldsymbol{0},\mathbf{M}_{n}), \quad
\mathbf{M}_{n} \in {\cal M}_{n},
\end{split}
\end{equation}
where the sample $\boldsymbol{\xi}_{n}^{T} = (\xi_{1},\ldots,\xi_{n})$
represents ``noise'' and consists of independent and identically
distributed (i.i.d.) Gaussian random variables with zero means and
variances $1$. The stochastic ``signal'' $\mathbf{\eta}_{n}$ is a
Gaussian random vector with zero mean and covariance matrix
$\mathbf{M}_{n}$. ${\cal M}_{n}$ is a given set of possible covariance
matrices $\mathbf{M}_{n}$.

Without loss of generality, we may assume a matrix ${\mathbf{M}}_{n}$
positive definite, i.e. $|{\mathbf{M}}_{n}|=\det{\mathbf{M}}_{n} > 0$. Indeed,
if $|{\mathbf{M}}_{n}|=0$, then measures ${\mathbf{P}}_{\boldsymbol{\xi}_{n}}$
and ${\mathbf{P}}_{\boldsymbol{\eta}_{n}}$ are orthogonal, and therefore
hypotheses $\mathcal H_{0}$ and $\mathcal H_{1}$ can be tested without errors.

In addition to model \eqref{mod1}, we consider also a similar model
\begin{equation}\label{mod2}
\begin{split}
&\mathcal H_{0}: {\mathbf y}_{n} = \boldsymbol{\xi}_{n}, \qquad
\boldsymbol{\xi}_{n} \sim
{\mathcal N}(\boldsymbol{0},\mathbf{I}_{n}), \\
&\mathcal H_{1}: {\mathbf y}_{n} = \boldsymbol{\xi}_{n} +
\mathbf{s}_{n}, \quad \mathbf{s}_{n} \sim
{\mathcal N}(\boldsymbol{0},\mathbf{S}_{n}), \quad
\mathbf{S}_{n} \in {\cal S}_{n},
\end{split}
\end{equation}
where $\mathbf{s}_{n}$ is a Gaussian vector independent of
$\boldsymbol{\xi}_{n}$ with
$\mathbf{s}_{n} \sim {\mathcal N}(\boldsymbol{0},\mathbf{S}_{n})$,
and ${\cal S}_{n}$ is a given set of possible covariance matrices
$\mathbf{S}_{n}.$ Model \eqref{mod2} is a particular
case of model \eqref{mod1}.

We now proceed with testing of hypotheses $\mathcal{H}_{0}$ and
$\mathcal{H}_{1}$ for model \eqref{mod1}. We select a decision region
${\mathcal{D}}\in\mathbf{R}^{n}$  such that
\begin{equation}\label{testD}
\begin{gathered}
{\mathbf y}_{n} \in \mathcal D \Rightarrow \mathcal H_{0}, \qquad
{\mathbf y}_{n} \not\in \mathcal D \Rightarrow \mathcal H_{1}.
\end{gathered}
\end{equation}
Then the 1-st kind error probability
(``{\it false alarm probability}'') $\alpha({\mathcal D})$ and the
2-nd kind error probability (``{\it miss probability}'')
$\beta({\mathcal D},{\cal M}_{n})$ are defined respectively, as
\begin{equation}\label{defalpha2}
\alpha({\mathcal D}) =
\mathbf{P}({\mathbf y}_{n} \not\in \mathcal D|\mathcal H_{0})
\end{equation}
and
\begin{equation}\label{defbeta2}
\begin{split}
\beta({\mathcal D},{\cal M}_{n}) &=
\mathbf{P}({\mathbf y}_{n} \in \mathcal D|\mathcal H_{1})  \\
&=\sup\limits_{\mathbf{M}_{n} \in {\cal M}_{n}}
\mathbf{P}({\mathbf y}_{n} \in \mathcal D|\mathbf{M}_{n}).
\end{split}
\end{equation}
Given the 1-st kind error probability $\alpha$, $0<\alpha<1,$ we
investigate the minimum possible 2-nd kind error probability
\begin{equation}\label{defbeta3a}
\beta(\alpha,{\cal M}_{n}) =
\inf\limits_{{\mathcal D}:\alpha({\mathcal D}) \leq \alpha}
\beta({\mathcal D},{\cal M}_{n})
\end{equation}
and the corresponding optimal decision region ${\mathcal D}(\alpha)$.

In this paper,  we consider the case when $\alpha$ is fixed (or slowly
vanishes with $n$). This case sometimes is called the Neyman-Pearson
minimax detection (or the Neyman-Pearson minimax testing of
hypotheses). In this case,  the 1-st kind and the 2-nd kind errors
imply very different losses for a statistician, and we are mainly
interested in minimization of the 2-nd kind
$\beta={\mathbf{P}}\{H_{0}|H_{1}\}$ error probability. The case is
quite popular in many applications (see, e.g., \cite{ZhangPoor11}
and references therein).

For a given $\mathbf{M}_{n}$ and a fixed $\alpha,$  let
$\beta(\mathbf{M}_{n})$ denote the minimum possible 2-nd kind error
probability. Similarly, for a given set
$\mathcal{M}_{n}=\{\mathbf{M}_{n}\}$ and a fixed $\alpha,$
$\beta(\mathcal{M}_{n})$ denotes the minimum of the minimax 2-nd
kind error probabilities (see \eqref{defbeta3a}). Clearly, we have
\begin{equation}\label{ineq1}
\begin{gathered}
\sup_{\mathbf{M}_{n} \in {\cal M}_{n}}\beta(\mathbf{M}_{n}) \leq
\beta({\cal M}_{n}),
\end{gathered}
\end{equation}
which is equivalent to
\begin{equation}\label{ineq1a}
\begin{gathered}
\sup_{\mathbf{M}_{n} \in {\cal M}_{n}}\inf_{{\mathcal D}}
\beta({\mathcal D},\mathbf{M}_{n}) \leq \inf_{{\mathcal D}}
\sup_{\mathbf{M}_{n} \in {\cal M}_{n}}
\beta({\mathcal D},{\cal M}_{n}).
\end{gathered}
\end{equation}

In many practical cases,  the value of $\beta(\mathbf{M}_{n})$
decreases exponentially as $n\rightarrow\infty$. Then,  it is
reasonable (e.g., simpler and more productive) to investigate the
exponential decay rates $n^{-1}\ln\beta(\mathbf{M}_{n})$ and
$n^{-1}\ln\beta(\mathcal{M}_{n})$ (some results on the equality in
\eqref{ineq1} can be found in \cite{Bur17a}).

For a fixed $\alpha$ and a given sequence of matrices $\mathbf{M}_{n},$ we
investigate sequences of sets $\mathcal{M}_{n}$, such that
$\mathbf{M}_{n} \in \mathcal{M}_{n}$ and the following equality holds
\begin{equation}\label{aseq1}
\lim_{n\to\infty}\frac{1}{n}\ln\beta(\mathbf{M}_{n}) =
\lim_{n\to\infty}\frac{1}{n}\ln\beta({\cal M}_{n}(\mathbf{M}_{n})).
\end{equation}
In other words, for a given 1-st kind error probability $\alpha$,
$\mathcal{M}_{n}$ is a set of covariance
matrices, which can be replaced by matrix $\mathbf{M}_{n}$. In the
sequel, we describe the maximal such sets $\mathcal{M}_{n}(\mathbf{M}_{n})$ for
both models \eqref{mod1} and \eqref{mod2} and  also give some
``inner bounds'' for them.

Motivation for considering minimax testing of hypotheses (detection
of signals) is described in detail in \cite{Wald, Lehmann, Poor}.
If relation \eqref{aseq1} holds for a given set of matrices
$\mathcal{M}_{n}$, then we may replace $\mathcal{M}_{n}$
(without asymptotic loss) by testing of a particular matrix
$\mathbf{M}_{n}$. Recall that the optimal test of a particular
matrix $\mathbf{M}_{n}$ is based on the Neyman-Pearson lemma, which
leads to a simple LRT ({\it Likelihood Ratio Test}) - detector.
Otherwise (without relation \eqref{aseq1}), the optimal minimax
test is only described by a much more complicated Bayes test with the
{\it least favorable} prior distribution on $\mathcal{M}_{n}$.
For this reason, it is natural to investigate
when a given set of matrices $\mathcal{M}_{n}$ can be replaced
by a particular matrix $\mathbf{M}_{n}$ \cite{ZhangPoor11}.
Technically, it is more convenient to consider an equivalent problem:
for a given matrix $\mathbf{M}_{n}$ find the largest set of
matrices $\mathcal{M}_{n}(\mathbf{M}_{n})$ that can be replaced
by the matrix $\mathbf{M}_{n}$. This problem is considered in
the sequel.

{\bf Definition 1}. For a fixed $\alpha$ and a given sequence of matrices
$\mathbf{M}_{n},$ define by $\mathcal{M}_{n}(\mathbf{M}_{n})$ the sequence
of the largest sets of matrices, such that
$\mathbf{M}_{n} \in \mathcal{M}_{n}(\mathbf{M}_{n})$, and
\begin{equation}\label{DEfM}
\begin{gathered}
\lim_{n\to\infty}\frac{1}{n}\ln\beta({\cal M}_{n}(\mathbf{M}_{n})) =
\lim_{n\to\infty}\frac{1}{n}\ln\beta(\mathbf{M}_{n}).
\end{gathered}
\end{equation}

In fact, it is convenient first to investigate similar to
$\mathcal{M}_{n}(\mathbf{M}_{n})$ the largest sets
$\mathcal{M}_{n}^{LR}(\mathbf{M}_{n})$, which arises if
LR-detector is used (see Definition 2 below). We will also show that
$\mathcal{M}_{n}(\mathbf{M}_{n})=\mathcal{M}_{n}^{LR}(\mathbf{M}_{n})$,
i.e. the LR-detector is asymptotically optimal.

In model \eqref{mod1}, denote by $\mathbf{P}_{\mathbf{I}_{n}}$ the distribution
of ${\mathbf y}_{n}= \boldsymbol{\xi}_{n}$, provided
$\boldsymbol{\xi}_{n} \sim {\mathcal N}(\boldsymbol{0},\mathbf{I}_{n})$.
Similarly, denote by $\mathbf{Q}_{\mathbf{M}_{n}}$ the distribution
of ${\mathbf y}_{n}= \boldsymbol{\eta}_{n}$, provided
$\boldsymbol{\eta}_{n} \sim {\mathcal N}(\boldsymbol{0},\mathbf{M}_{n})$.
Also, denote by $p_{\mathbf{I}_{n}}({\mathbf y}_{n})$,
${\mathbf y}_{n} \in {\mathbf R}^{n}$ and $p_{\mathbf{M}_{n}}({\mathbf y}_{n})$
the corresponding probability density functions.

Note that if $\left|\mathbf{V}_{n}\right| \neq 0$ and
$\left|\mathbf{M}_{n}\right| \neq 0$, then
\begin{equation}\label{deGaus2}
\begin{split}
\ln\frac{p_{\mathbf{V}_{n}}}{p_{\mathbf{M}_{n}}}({\mathbf y}_{n}) = \frac{1}{2}
\left[\ln\frac{\left|\mathbf{M}_{n}\right|}{\left|\mathbf{V}_{n}\right|} +
\left({\mathbf y}_{n},\left(\mathbf{M}_{n}^{-1} - \mathbf{V}_{n}^{-1}\right)
{\mathbf y}_{n}\right)\right].
\end{split}
\end{equation}
If $\mathbf{V}_{n} = \mathbf{I}_{n}$, then
\begin{equation}\label{deGaus3}
\begin{gathered}
\ln\frac{p_{\mathbf{I}_{n}}}{p_{\mathbf{M}_{n}}}({\mathbf y}_{n}) =
\frac{1}{2}\left[\ln\left|\mathbf{M}_{n}\right| +
\left({\mathbf y}_{n},\left(\mathbf{M}_{n}^{-1} - \mathbf{I}_{n}\right)
{\mathbf y}_{n}\right)\right].
\end{gathered}
\end{equation}
Introduce the logarithm of the likelihood ration (see \eqref{deGaus3})
\begin{equation}\label{LLR1}
\begin{split}
&f_{\mathbf{M}_{n}}\left({\mathbf y}_{n}\right) =
\ln\frac{p_{\mathbf{I}_{n}}}{p_{\mathbf{M}_{n}}}({\mathbf y}_{n}) = 
\frac{1}{2}\left[\ln\left|\mathbf{M}_{n}\right| + \left({\mathbf y}_{n},
\left(\mathbf{M}_{n}^{-1} -\mathbf{I}_{n}\right){\mathbf y}_{n}\right)\right].
\end{split}
\end{equation}

We consider LR-detector with corresponding decision regions.
Introduce the decision region ${\cal D}_{LR}(\mathbf{M}_{n},\alpha)$ in
favor of $\mathbf{I}_{n}$ when testing the matrices $\mathbf{I}_{n}$ and
$\mathcal{M}_{n}$:
\begin{equation}\label{DLR}
\begin{split}
&{\cal D}_{LR}(\mathbf{M}_{n},\alpha) =
\left\{\mathbf y_{n} \in \mathbf{R}^{n}: f_{\mathbf{M}_{n}}
\left({\mathbf y}_{n}\right) \geq \gamma\right\},
\end{split}
\end{equation}
where $\gamma$ is such that
\begin{equation}\label{DLR1}
\begin{split}
&\alpha = \mathbf{P}_{\mathbf{I}_{n}}\left\{
{\cal D}_{LR}^{c}(\mathbf{M}_{n},\alpha)\right\} =
\mathbf{P}_{\mathbf{I}_{n}}\left\{f_{\mathbf{M}_{n}}
\left({\mathbf y}_{n}\right) \leq \gamma\right\} =\\
&= \mathbf{P}_{\mathbf{I}_{n}}\left\{
\left[\left(\boldsymbol{\xi}_{n},
\left(\mathbf{M}_{n}^{-1} -\mathbf{I}_{n}\right)\boldsymbol{\xi}_{n}
\right) + \ln\left|\mathbf{M}_{n}\right|\right] \leq 2\gamma\right\}.
\end{split}
\end{equation}

In model \eqref{mod1}, assume that for testing matrices ${\mathbf{I}}_{n}$ and
${\mathbf{M}}_{n}$ (i.e. simple hypotheses) we use the optimal detector
(i.e. LRT-detector) with the decision region
${\cal D}_{LR}({\mathbf{M}}_{n},\alpha)$ (see \eqref{DLR}-\eqref{DLR1})
in favor of $\mathbf{I}_{n}$. For what matrices ${\mathbf{V}}_{n}$ instead of
${\mathbf{M}}_{n}$ the decision region
${\cal D}_{LR}({\mathbf{M}}_{n},\alpha)$ does not deteriorate the 2-nd
kind error probability $\beta(\alpha,{\mathbf{M}}_{n})$ ? In order to answer
that question, introduce

{\bf Definition 2}. For a fixed $\alpha$ and a given sequence of matrices
$\mathbf{M}_{n},$ define by \\
$\mathcal{M}_{n}^{LR}(\mathbf{M}_{n})$ the sequence
of the largest sets of matrices $\mathbf{V}_{n},$ such that
\begin{equation}\label{aseq1a}
\begin{gathered}
\lim_{n\to\infty}\frac{1}{n}\ln
\sup_{\mathbf{V}_{n} \in \mathcal{M}_{n}^{LR}(\mathbf{M}_{n})}
\beta(\mathbf{V}_{n}) \leq \lim_{n\to\infty}\frac{1}{n}\ln\beta(\mathbf{M}_{n}),
\end{gathered}
\end{equation}
provided the decision regions
${\cal D}_{LR}(\alpha,{\mathbf{M}}_{n})$ are used.

\subsection{Kullback--Leibler distance}
For the random elements ${\mathbf{x}}$ and ${\mathbf{y}}$ defined on a
measurable space $(\Omega,\mathcal{B})$ with probability distributions
${\mathbf{P}}_{\mathbf{x}}$ and ${\mathbf{Q}}_{\mathbf{y}}$, respectively,
introduce the function
\begin{equation}\label{Stein0}
\begin{gathered}
D({\mathbf{P}}_{\mathbf{x}}||{\mathbf{Q}}_{\mathbf{y}}) =
{\mathbf E}_{{\mathbf{P}}_{\mathbf{x}}}
\ln\frac{d{\mathbf P}_{\mathbf x}}
{d{\mathbf Q}_{\mathbf y}}\left({\mathbf u}\right),
\end{gathered}
\end{equation}
(\emph{Kullback--Leibler distance} or \emph{divergence} for measures
${\mathbf{P}}_{\mathbf{x}}$ and ${\mathbf{Q}}_{\mathbf{y}}$).

In particular, if ${\mathbf x},{\mathbf y} \in {\mathbf R}^{n}$, and
${\mathbf x} \sim
{\mathcal N}(\boldsymbol{0},\mathbf{V}_{n})$,
${\mathbf y} \sim {\mathcal N}(\boldsymbol{0},\mathbf{M}_{n})$, then
\cite[Ch. 9.1]{Kullback} (${\rm tr}$ = trace)
\begin{equation}\label{Stein1}
\begin{gathered}
D({\mathbf{P}}_{\mathbf{x}}||{\mathbf{Q}}_{\mathbf{y}}) =
\frac{1}{2}\ln\frac{\left|\mathbf{M}_{n}\right|}{\left|\mathbf{V}_{n}\right|}
+ \frac{1}{2}{\rm tr}\left(\mathbf{V}_{n}\mathbf{M}_{n}^{-1}\right) -
\frac{n}{2}.
\end{gathered}
\end{equation}
If $\mathbf{V}_{n} = \mathbf{I}_{n}$, then
\begin{equation}\label{Stein2}
\begin{split}
D(\mathbf{I}_{n}||\mathbf{M}_{n}) =
&D({\mathbf{P}}_{\mathbf{x}}||{\mathbf{Q}}_{\mathbf{y}}) = 
\frac{1}{2}\sum_{i=1}^{n}\left(\ln\lambda_{i} +
\frac{1}{\lambda_{i}}-1\right),
\end{split}
\end{equation}
where $\lambda_{1},\ldots,\lambda_{n}$ are eigenvalues of the matrix
$\mathbf{M}_{n}$ (eigenvalues of the matrix $\mathbf{M}_{n}^{-1}$ are
$\lambda_{1}^{-1},\ldots,\lambda_{n}^{-1}$).

Kullback--Leibler distance plays important role in testing of hypotheses.
For example, in model \eqref{mod1}, assume that for testing matrices
${\mathbf{I}}_{n}$ and ${\mathbf{M}}_{n}$ (i.e. simple hypotheses) we use the
optimal detector, that is the LRT-detector with the decision region
${\cal D}_{LR}(\alpha,{\mathbf{M}}_{n})$ (see \eqref{DLR}-\eqref{DLR1}) in favor
of $\mathbf{I}_{n}$. Then under some natural assumptions the following formula
holds
\begin{equation}\label{Stein01}
\begin{gathered}
\lim_{\alpha\to 0}\lim_{n\to \infty}\frac{1}{n}\ln\beta(\alpha) =
-\lim_{n\to \infty}\frac{1}{n}D(\mathbf{I}_{n}||\mathbf{M}_{n}).
\end{gathered}
\end{equation}

Relation \eqref{Stein01} is called Stein's lemma \cite{Stein, Chernov56}. In
the case of independent
identically distributed random variables, its proof can be found in
\cite[Theorem 3.3]{Kullback}, \cite[Theorem 12.8.1]{CT}. It is natural to expect
that formula \eqref{Stein01} holds not only for testing simple hypotheses,
but in more general cases of testing composite hypotheses. Some particular
analogs of formula \eqref{Stein01} have already appeared for the cases of
stationary Gaussian \cite{ZhangPoor11} and Poisson \cite{Bur21} random processes.

In this paper, some analogs of formula \eqref{Stein01} for models \eqref{mod1}
and \eqref{mod2} are derived.

\subsection{Assumptions}
Let $\mathcal{C}_{n}$ be the convex set of all $n\times n$ - covariance (i.e.
positive definite symmetric) matrices in $\mathbf{R}^{n}$.
For model \eqref{mod1}, we consider a sequence of sets
${\cal M}_{i} \in \mathcal{C}_{i}$
of covariance matrices $\mathbf{M}_{i} \in {\cal M}_{i}$,
$i=1,2,\ldots$, in a ``scheme of series'', e.g. ${\cal M}_{i+1}$
is not necessarily a ``continuation'' of ${\cal M}_{i}$.
We denote by
$\lambda_{1}(\mathbf{M}_{n}),\ldots,\lambda_{n}(\mathbf{M}_{n})$
the eigenvalues (all positive) of the covariance matrix
$\mathbf{M}_{n}$. We assume that the following assumptions are
satisfied:

{\bf I}. For all matrices $\mathbf{M}_{n} \in {\cal M}_{n}$
there exist positive limits as $n\to \infty$ (see \eqref{Stein1a})
\begin{equation}\label{assump0}
\begin{gathered}
\lim_{n\to\infty}\frac{1}{n}\sum_{i=1}^{n}
\left(\ln\lambda_{i}(\mathbf{M}_{n}) +
\frac{1}{\lambda_{i}(\mathbf{M}_{n})}-1\right),
\end{gathered}
\end{equation}
where convergence is uniform on $\mathbf{M}_{n} \in {\cal M}_{n}$
(note that $\ln z \geq 1-1/z$, $z > 0$).

{\bf II}. For some $\delta > 0$ we have
\begin{equation}\label{assump3}
\begin{gathered}
\lim_{n\to\infty}\frac{1}{n}\sup_{\mathbf{M}_{n} \in {\cal M}_{n}}
\sum_{i=1}^{n}\left|\frac{1}{\lambda_{i}(\mathbf{M}_{n})}-1
\right|^{1+\delta} < \infty.
\end{gathered}
\end{equation}

\subsection{Main results}
In this paper, for a $n\times n$ matrix ${\mathbf{A}}_{n}$, we use
notation $|{\mathbf{A}}_{n}|=\det{\mathbf{A}}_{n}$. Also, let
$({\mathbf{x}},{\mathbf{y}})$ denote the inner product of two vectors
${\mathbf{x}},{\mathbf{y}}$.  We write ${\mathbf{A}}_{n}>0$  if
the matrix ${\mathbf{A}}_{n}$ is positive definite.

For any $\mathbf{M}_{n},{\mathbf{V}}_{n}\in\mathcal{C}_{n},$ such that
$\mathbf{I}_{n}+\mathbf{V}_{n}^{-1} -\mathbf{M}_{n}^{-1} > 0$,
define the function
\begin{equation}\label{Theor1a}
\begin{split}
f(\mathbf{M}_{n},{\mathbf V}_{n}) = \frac{\left|\mathbf{M}_{n}\right|^{1/2}}
{\left|\mathbf{I}_{n}+
\mathbf{V}_{n}\left(\mathbf{I}_{n}-\mathbf{M}_{n}^{-1}\right)\right|^{1/2}}.
\end{split}
\end{equation}
Note that 
\begin{equation}\label{Theor1ab}
\begin{split}
\mathbf{I}_{n}+\mathbf{V}_{n}\left(\mathbf{I}_{n}-\mathbf{M}_{n}^{-1}\right) =
\mathbf{M}_{n}+(\mathbf{V}_{n}-\mathbf{M}_{n})
\left(\mathbf{I}_{n}-\mathbf{M}_{n}^{-1}\right).
\end{split}
\end{equation}

The main result of the paper describes the sets
$\mathcal{M}_{n}(\mathbf{M}_{n})$ and
$\mathcal{M}_{n}^{LR}(\mathbf{M}_{n})$.

{\bf Theorem 1}. \textit{Let assumptions \eqref{assump0}-\eqref{assump3}
hold for model \eqref{mod1}. Then (as $n\to \infty$)
\begin{equation}\label{Theor1}
\begin{split}
&\mathcal{M}_{n}(\mathbf{M}_{n}) =
\mathcal{M}_{n}^{LR}(\mathbf{M}_{n}) = 
\left\{\mathbf{V}_{n}\in {\cal C}_{n}:
\sup\limits_{\mathbf{V}_{n} \in \mathcal{M}_{n}(\mathbf{M}_{n})}
{\mathbf E}_{\mathbf{I}_{n}}\frac{p_{\mathbf{V}_{n}}}
{p_{\mathbf{M}_{n}}}({\mathbf x}) \leq e^{o(n)}\right\}  \\
&= \left\{\mathbf{V}_{n}:
\begin{array}{c}
\mathbf{I}_{n}+\mathbf{V}_{n}^{-1} -\mathbf{M}_{n}^{-1} > 0, \\
\sup\limits_{\mathbf{V}_{n} \in \mathcal{M}_{n}(\mathbf{M}_{n})}
f(\mathbf{M}_{n},{\mathbf V}_{n}) \leq e^{o(n)}
\end{array}
\right\},
\end{split}
\end{equation}
where the function $f(\mathbf{M}_{n},{\mathbf V}_{n})$ is defined in
\eqref{Theor1a}.}

Clearly, the sets $\mathcal{M}_{n}(\mathbf{M}_{n})$ and
$\mathcal{M}_{n}^{LR}(\mathbf{M}_{n})$ are convex.

{\it Remark 1}. It is known \cite[Theorem 7.6.7]{Horn},
\cite[Ch. 8.5, Theorem 4]{Bellman} that the function
$f({\mathbf A}_{n})= \ln|{\mathbf A}_{n}|$ is strictly concave
on the convex set $\mathcal{C}_{n}$ of positive definite symmetric matrices
in $\mathbf{R}^{n}$. From that result also follows convexity of
the set $\mathcal{M}_{n}(\mathbf{M}_{n})$, i.e. if
$\mathbf{V}_{n}^{(1)} \in\mathcal{M}_{n}(\mathbf{M}_{n})$ and
$\mathbf{V}_{n}^{(2)}\in \mathcal{M}_{n}(\mathbf{M}_{n})$, then
$a\mathbf{V}_{n}^{(1)}+(1-a)\mathbf{V}_{n}^{(2)} \in
\mathcal{M}_{n}(\mathbf{M}_{n})$ for any $0\leq a\leq1$.

We present also a simplified consequence to Theorem 1, limiting ourselves in
\eqref{Theor1} only to matrices $\mathbf{V}_{n}$, commutating with
$\mathbf{M}_{n}$. For a matrix $\mathbf{M}_{n}\in {\cal C}_{n}$ introduce
the convex set ${\cal C}_{\mathbf{M}_{n}}$ of covariance matrices
$\mathbf{V}_{n}$, commutating with $\mathbf{M}_{n}$:
\begin{equation}\label{defC}
\begin{split}
{\cal C}_{\mathbf{M}_{n}} =\left\{\mathbf{V}_{n}\in {\cal C}_{n}:
\mathbf{M}_{n}\mathbf{V}_{n} =\mathbf{V}_{n}\mathbf{M}_{n}\right\}.
\end{split}
\end{equation}
Denote by $\{\lambda_{i}\}$ the eigenvalues of $\mathbf{M}_{n}$ and by
$\{\nu_{i}\}$ the eigenvalues of $\mathbf{V}_{n}$. Then the function
$f(\mathbf{M}_{n},{\mathbf V}_{n})$ from \eqref{Theor1a} takes the form
\begin{equation}\label{deffVM}
\begin{split}
f(\mathbf{M}_{n},{\mathbf V}_{n}) =
\prod_{i=1}^{n}\frac{\lambda_{i}}
{[\lambda_{i}+\nu_{i}(\lambda_{i}-1)]^{1/2}},
\end{split}
\end{equation}
provided $\lambda_{i}+\nu_{i}(\lambda_{i}-1) > 0$, $i=1,\ldots,n$.

Introduce the following subset of $\mathcal{C}_{\mathbf{M}_{n}}$
as $n \to \infty$:
\begin{equation}\label{defV0}
\begin{split}
&{\cal V}_{n}^{(0)}(\mathbf{M}_{n}) = 
\left\{{\mathbf V}_{n}:
\sup\limits_{\mathbf{V}_{n} \in {\cal V}_{n}^{(0)}(\mathbf{M}_{n})}
f(\mathbf{M}_{n},{\mathbf V}_{n}) \geq e^{o(n)}\right\},
\end{split}
\end{equation}
where the function $f(\mathbf{M}_{n},{\mathbf V}_{n})$ is defined in
\eqref{deffVM}. Then the following ``inner bound''
${\cal V}_{n}^{(0)}(\mathbf{M}_{n})$ for ${\cal M}_{n}(\mathbf{M}_{n})$ holds.

{\bf Theorem 2}.
\textit{Let assumptions \eqref{assump0}-\eqref{assump3} hold for model
\eqref{mod1}. Then  the set $\mathcal{M}_{n}(\mathbf{M}_{n})$ contains the set
$\mathcal{V}_{n}^{(0)}(\mathbf{M}_{n})$:}
\begin{equation}\label{Cor1}
\begin{gathered}
{\cal V}_{n}^{(0)}(\mathbf{M}_{n}) \subseteq
{\cal M}_{n}(\mathbf{M}_{n}).
\end{gathered}
\end{equation}

The set ${\cal V}_{n}^{(0)}(\mathbf{M}_{n})$ is convex in ${\mathbf V}_{n}$
(see Remark 1).

{\it Remark 2}. In the right-hand side of \eqref{defV0}, replacing
$o(n)$ by  $0$, consider the set
\begin{equation}\label{defV1}
\begin{split}
{\cal V}_{n}^{(1)}(\mathbf{M}_{n}) =\left\{{\mathbf V}_{n}:
\sup\limits_{\mathbf{V}_{n} \in {\cal V}_{n}^{(1)}(\mathbf{M}_{n})}
f(\mathbf{M}_{n},{\mathbf V}_{n}) \leq 1\right\}.
\end{split}
\end{equation}
Clearly, $\mathcal{V}_{n}^{(1)}(\mathbf{M}_{n})\in\mathcal{V}_{n}^{(0)}
(\mathbf{M}_{n})$. In a sense, the set
$\mathcal{V}_{n}^{(0)}(\mathbf{M}_{n})$ is the set
$\mathcal{V}_{n}^{(1)}(\mathbf{M}_{n})$, enlarged by a
``thin slice'' whose width has the order of $o(n)$. In other words,
$\mathcal{V}_{n}^{(1)}(\mathbf{M}_{n})$ can be considered as a
``core'' of the set $\mathcal{V}_{n}^{(0)}(\mathbf{M}_{n})$.

This paper is inspired by paper \cite{ZhangPoor11}, where a
similar minimax detection problem for stochastic stationary signals
was considered. We consider a more general case of the Gaussian vectors
with unknown covariance matrices, which in turn yields a  more
natural and convenient way to proceed with the particular case of
the stationary stochastic signals.

The approach of this paper was used earlier in \cite{Bur21} for the
case of the Poisson processes. Some other important cases with
additional constraints on error probabilities $\alpha,\beta$ were
considered in \cite{Bur79, Bur17}.

\subsection{Inverse problem}

Following paper \cite{ZhangPoor11}, we consider also the inverse problem.
It corresponds to the following question: when the testing of a given
set of matrices $\mathcal{M}_{n}$ can be replaced by the testing of some matrix
$\mathbf{M}_{n}^{(0)}$ ?

A sufficient condition to such replacement follows from Theorem 1. Let a
sequence of sets $\{\mathcal{M}_{1},\mathcal{M}_{2},\ldots\}$,
$\mathcal{M}_{i}\subset\mathcal{C}_{i}$, $i=1,2,\ldots$ of covariance
matrices $\mathbf{M}_{i}\in\mathcal{M}_{i}$ be given. Denote by
$\{\mathbf{M}_{i}^{(0)}\}$ a sequence of covariance matrices
$\mathbf{M}_{i}^{(0)}$ (if exists)
that satisfies the following analog of  relation \eqref{aseq1}
\begin{equation}\label{aseq3}
\begin{split}
\lim_{n\to\infty}\frac{1}{n}\ln\beta(\mathbf{M}_{n}^{(0)}) =
&\lim_{n\to\infty}\frac{1}{n}\ln\beta({\cal M}_{n}) =
\lim_{n\to\infty}\frac{1}{n}
\ln\beta(\{{\cal M}_{n},\mathbf{M}_{n}^{(0)}\}),
\end{split}
\end{equation}
where $\{{\cal M}_{n},\mathbf{M}_{n}^{(0)}\} =
{\cal M}_{n}\bigcup \mathbf{M}_{n}^{(0)}$ is the ``enlargement'' of
${\cal M}_{n}$ by $\mathbf{M}_{n}^{(0)}$. We do not require that
$\mathbf{M}_{i}^{(0)} \in {\cal M}_{i}$. When there exists a sequence
$\{\mathbf{M}_{i}^{(0)}\}$, satisfying \eqref{aseq3} ?

As a corollary to Theorem 1, we obtain a sufficient condition for having
\eqref{aseq3}:

{\bf Proposition 1}.
\textit{For model \eqref{mod1}, let $\{\mathcal{M}_{i}\}$ be an
arbitrary sequence satisfying \\
assumptions \eqref{assump0}-\eqref{assump3}.
If for a sequence $\{\mathbf{M}_{i}^{(0)}\}$ the following conditions are
fulfilled: \\
1) $\mathbf{I}_{n}+\mathbf{V}_{n}^{-1} -\left(\mathbf{M}_{n}^{(0)}\right)^{-1}
> 0$  \ for all $\mathbf{V}_{n} \in {\cal M}_{n}$; \\
2)
\begin{equation}\label{Theor3}
\begin{split}
\sup_{\mathbf{V}_{n} \in {\cal M}_{n}}
f(\mathbf{M}_{n}^{(0)},{\mathbf V}_{n}) \leq e^{o(n)}, \quad n \to \infty,
\end{split}
\end{equation}
then the sequence $\{\mathbf{M}_{i}^{(0)}\}$ together with LRT-detectors
satisfies  condition \eqref{aseq3}.}

{\it Remark 3}. If  a sequence $\{\mathbf{M}_{i}^{(0)}\}$ satisfies
condition \eqref{Theor3} for a sequence of sets
$\{\mathcal{M}_{i}\}$, then it will also satisfy that condition
for the sequence of sets $\{{\rm conv}\mathcal{M}_{i}\}$, where
${\rm conv}\mathcal{M}_{i}$ is the smallest convex set of
matrices, containing the set $\mathcal{M}_{i}$. Clearly,
$\mathcal{M}_{i} \subseteq {\rm conv}\mathcal{M}_{i}$.
The set ${\rm conv}\mathcal{M}_{i}$ sometimes is called
``convex hull'' of $\mathcal{M}_{i}$.

Proposition 1 generalizes  a similar result of
\cite[Theorem 1]{ZhangPoor11} (see Corollary 3 in Section III.B).

The paper is organized as follows. In Section II, we present and prove
quite important for us  auxiliary Theorem 3. In Section III we prove
Theorems 1 and 2 along with some related results. Model \eqref{mod2}
and the case of the stationary stochastic signals are considered
in Section IV. In essence, all results of  Section IV represent
the corollaries of Theorem 2 for model \eqref{mod1}. Some
applications of our results are presented in Section V as
specific examples.

\section{Auxiliary Theorem 3 with Proof}
Note that model \eqref{mod1} can be reduced to the equivalent case with
a diagonal matrix $\mathbf{M}_{n}$. Indeed, since $\mathbf{M}_{n}$ is
a covariance matrix (i.e. symmetric and nonnegative definite),
there is an orthogonal matrix $\mathbf{T}_{n}$ and a diagonal
matrix $\mathbf{\Lambda}_{n}$ such that $\mathbf{M}_{n} =
\mathbf{T}_{n}\mathbf{\Lambda}_{n}\mathbf{T}_{n}'$
\cite[Ch. 4.7-9]{Bellman}, \cite[Theorem 4.1.5]{Horn}.
The diagonal matrix
$\mathbf{\Lambda}_{n}= \mathbf{T}_{n}'\mathbf{M}_{n}\mathbf{T}_{n}$
consists of the eigenvalues $\{\lambda_{i}\}$ of $\mathbf{M}_{n}$.

Note also that for any orthogonal matrix $\mathbf{T}_{n},$  a vector
$\mathbf{T}_{n}^{\prime}\boldsymbol{\xi}_{n}$ has the same
distribution as that of $\boldsymbol{\xi}_{n}$ (for the simple
hypothesis $\mathcal{H}_{0}$ of \eqref{mod1}).  Therefore,
multiplying both sides of \eqref{mod1} by $\mathbf{T}_{n}^{\prime}$,
we may reduce model \eqref{mod1} to the equivalent case with a
diagonal matrix $\mathbf{M}_{n}$.

\subsection{Simple Hypotheses}

In model \eqref{mod1}, we first consider the testing of matrices
${\mathbf{I}}_{n}$ and ${\mathbf{M}}_{n}$ (i.e. simple hypotheses), using
the optimal detector. Denote
$D(\mathbf{I}_{n}||\mathbf{M}_{n}) =
D({\mathbf P}_{\mathbf{I}_{n}}||{\mathbf Q}_{\mathbf{M}_{n}})$.

The main auxiliary result of this paper is the following.

{\bf Theorem 3}. \textit{
The minimal possible $\beta(\alpha)$,
$0 < \alpha<1$, satisfies the bounds
\begin{equation}\label{Stein11}
\begin{gathered}
\ln\beta(\alpha)\geq -\frac{D(\mathbf{I}_{n}||\mathbf{M}_{n})+
h(\alpha)}{1-\alpha},\\
h(\alpha) = -\alpha\ln\alpha - (1-\alpha)\ln(1-\alpha),
\end{gathered}
\end{equation}
and
\begin{equation}\label{Stein12}
\begin{gathered}
\ln\beta(\alpha) \leq -D(\mathbf{I}_{n}||\mathbf{M}_{n}) +
\mu_{0}(\alpha,\mathbf{M}_{n}),
\end{gathered}
\end{equation}
where $\mu_{0}(\alpha,\mathbf{M}_{n})$ is defined by the relation}
\begin{equation}\label{defmu0}
\begin{gathered}
{\mathbf P}_{\mathbf{I}_{n}}\left\{\ln\frac{p_{\mathbf{I}_{n}}}
{p_{\mathbf{M}_{n}}}({\mathbf x}) \leq
D(\mathbf{I}_{n}||\mathbf{M}_{n})-\mu_{0}\right\} = \alpha.
\end{gathered}
\end{equation}

Note that bounds \eqref{Stein11} and \eqref{Stein12} are pure analytical
relations without any limiting operations. Also, both lower bound
\eqref{Stein11} and upper bound \eqref{Stein12} are close to each other,
if the value $\mu_{0}(\alpha,\mathbf{M}_{n})$ is much smaller than
$D(\mathbf{I}_{n}||\mathbf{M}_{n})$ (which usually has the order of $n$).

Next result gives an upper bound for $\mu_{0}(\alpha,\mathbf{M}_{n})$ of the
order $n^{1/p}$, $p> 1$ (see proof in Appendix).

{\bf Corollary 1}. \textit{Assume that the following condition is fulfilled
for some $1 < p \leq 2$
\begin{equation}\label{assump3b}
\begin{gathered}
\sup_{n}\frac{1}{n}
\sum_{i=1}^{n}\left|\frac{1}{\lambda_{i}(\mathbf{M}_{n})}-1
\right|^{p} \leq C_{p} < \infty.
\end{gathered}
\end{equation}
Then for $\mu_{0}(\alpha,\mathbf{M}_{n})$ from \eqref{Stein12} the upper
bound holds}
\begin{equation}\label{lem1a}
\begin{gathered}
\mu_{0}(\alpha,\mathbf{M}_{n}) \leq \left(\frac{3C_{p}n}{\alpha}\right)^{1/p}.
\end{gathered}
\end{equation}

In particular, if condition \eqref{assump3b} is fulfilled for $p= 2$,
then \eqref{lem1a} gives for $\mu_{0}(\alpha,\mathbf{M}_{n})$ the upper bound
of the order $\sqrt{n/\alpha}$. Below, Corollary 1 will be used together with
the assumption \eqref{assump3}.

\subsection{Proof of Theorem 3}
We first derive lower bound \eqref{Stein11}. Let ${\cal D} \in \mathbf{R}^{n}$
be a decision region in favor of $\mathbf{I}_{n}$, and $\beta=\beta({\cal D})$,
$\alpha = \alpha({\cal D})$ be the corresponding error probabilities.
Then denoting $p=p_{\mathbf{I}_{n}}$ and $q=p_{\mathbf{M}_{n}}$,
we have with ${\cal D}^{c} = \mathbf{R}^{n} \setminus {\cal D}$
\begin{equation}\label{Defab}
\begin{split}
&\beta = {\mathbf Q}_{\mathbf{M}_{n}}({\cal D}) = \int\limits_{{\cal D}}
p({\mathbf x})\frac{q}{p}({\mathbf x})d{\mathbf x},  \qquad
\alpha = {\mathbf P}_{\mathbf{I}_{n}}({\cal D}^{c}).
\end{split}
\end{equation}
Since ${\mathbf P}_{\mathbf{I}_{n}}({\cal D})=1-\alpha$, then considering
${\mathbf P}_{\mathbf{I}_{n}}/(1-\alpha)$ as the probability distribution on
${\cal D}$, and using the inequality
$\ln {\mathbf E}\xi \geq {\mathbf E}\ln \xi$, we have
\begin{equation}\label{Stein1a}
\begin{split}
&\ln\frac{\beta}{1-\alpha} = \ln\left[\frac{1}{(1-\alpha)}
\int\limits_{{\cal D}}p({\mathbf x})\frac{q}{p}({\mathbf x})d{\mathbf x}
\right]  \geq \frac{1}{(1-\alpha)}
\int\limits_{{\cal D}}p({\mathbf x})\ln\frac{q}{p}({\mathbf x})d{\mathbf x}  \\
&= - \frac{D({\mathbf I}_{n}||{\mathbf M}_{n})}{1-\alpha} -
\frac{1}{(1-\alpha)}
\int\limits_{{\cal D}^{c}}p({\mathbf x})\ln\frac{q}{p}({\mathbf x})d{\mathbf x}.
\end{split}
\end{equation}
Since ${\mathbf P}_{\mathbf{I}_{n}}({\cal D}^{c})=\alpha$, similarly to
\eqref{Stein1a}, the last term in the right-hand side of \eqref{Stein1a} gives
\begin{equation}\label{Stein1b}
\begin{split}
\int\limits_{{\cal D}^{c}}p({\mathbf x})
\ln\frac{q}{p}({\mathbf x})d{\mathbf x}
\leq \alpha\ln\left[\frac{1}{\alpha}
\int\limits_{{\cal D}^{c}}q({\mathbf x})d{\mathbf x}\right] = 
\alpha\ln\frac{1-\beta}{\alpha}
\leq \alpha\ln\frac{1}{\alpha}.
\end{split}
\end{equation}
Therefore, from \eqref{Stein1a} and \eqref{Stein1b} we have
\begin{equation}\label{Stein1c}
\begin{gathered}
\ln\frac{\beta}{1-\alpha} \geq
- \frac{D({\mathbf I}_{n}||{\mathbf M}_{n})}{1-\alpha} -
\frac{\alpha}{(1-\alpha)}\ln\frac{1}{\alpha},
\end{gathered}
\end{equation}
from where lower bound \eqref{Stein11} follows.

In order to prove upper bound \eqref{Stein12}, we set a value
$\mu > 0$, and define the acceptance region in favor of $\mathbf{I}_{n}$
\begin{equation}\label{SteinP1}
\begin{gathered}
{\cal A}_{\mu}= \left\{{\mathbf x} \in {\cal X}:\ln\frac{p}{q}({\mathbf x})
\geq D({\mathbf I}_{n}||{\mathbf M}_{n})-\mu\right\}.
\end{gathered}
\end{equation}
Denote by $\alpha_{\mu}$ and $\beta_{\mu}$ the first and
the second kind error probabilities for the acceptance region
${\cal A}_{\mu}$, respectively. Then by \eqref{SteinP1}
\begin{equation}\label{SteinQ1}
\begin{split}
\beta_{\mu} = \int\limits_{{\cal A}_{\mu}}p({\mathbf x})
\frac{q}{p}({\mathbf x})d{\mathbf x} =
e^{-D({\mathbf I}_{n}||{\mathbf M}_{n})+\mu_{1}},
\end{split}
\end{equation}
where $0 \leq \mu_{1} \leq \mu$. Also
\begin{equation}\label{SteinP1a}
\begin{split}
\alpha_{\mu} = {\mathbf P}_{{\mathbf I}_{n}}\left\{
\ln\frac{p}{q}({\mathbf x}) \leq
D({\mathbf I}_{n}||{\mathbf M}_{n})-\mu\right\} =
{\mathbf P}(\eta \geq \mu),
\end{split}
\end{equation}
where
\begin{equation}\label{defeta}
\eta = D({\mathbf I}_{n}||{\mathbf M}_{n}) - \ln\frac{p}{q}({\mathbf x}),
\qquad {\mathbf E}_{{\mathbf I}_{n}}\eta =0.
\end{equation}
Best is to set $\mu$ such that $\alpha_{\mu}=\alpha$. Therefore, we define
$\mu=\mu_{0}(\alpha,\mathbf{M}_{n})$ by formula \eqref{defmu0}. Then we have
\begin{equation}\label{SteinQ3b}
\begin{gathered}
\alpha_{\mu_{0}(\alpha,\mathbf{M}_{n})} = \alpha,
\end{gathered}
\end{equation}
and by \eqref{SteinQ1}
\begin{equation}\label{SteinQ2b}
\begin{gathered}
\ln\beta_{\mu_{0}(\alpha,\mathbf{M}_{n})} \leq
-D({\mathbf I}_{n}||{\mathbf M}_{n}) + \mu_{0}(\alpha,\mathbf{M}_{n}),
\end{gathered}
\end{equation}
which proves upper bound \eqref{Stein12}.  \qquad $\Box$

\section{Proofs of Theorem 1 and Theorem 2}
Since $\mathcal{M}_{n}^{LR}(\mathbf{M}_{n}) \subseteq
\mathcal{M}_{n}(\mathbf{M}_{n})$, in order to prove Theorem 1 it is sufficient
to get the ``inner bound'' for $\mathcal{M}_{n}^{LR}(\mathbf{M}_{n})$, and then
to get a similar ``outer bound'' for $\mathcal{M}_{n}(\mathbf{M}_{n})$.

\subsection{``Inner bound''}
We begin with the ``inner bound'' for $\mathcal{M}_{n}^{LR}(\mathbf{M}_{n})$.
Consider the testing of the simple hypothesis ${\mathbf{I}}_{n}$ against a
composite alternative $\mathcal M_{n}$. We use the optimal LRT-detector for a
matrix $\mathbf{M}_{n}\in \mathcal M_{n}$, with the decision regions
${\cal D}_{LR}(\mathbf{M}_{n},\alpha)={\cal A}_{\mu_{0}}$ in favor of
$\mathbf{I}_{n}$ (see \eqref{DLR}-\eqref{DLR1} and \eqref{SteinP1}), where
$\mu_{0} =\mu_{0}(\alpha,\mathbf{M}_{n})> 0$ is defined in \eqref{defmu0}.
Consider an another matrix $\mathbf{V}_{n}\in \mathcal M_{n}$, and evaluate
the 2-nd kind error probability $\beta(\alpha,{\mathbf{V}}_{n})$, provided
the decision regions ${\cal A}_{\mu_{0}}$ are used. By
\eqref{SteinP1}-\eqref{SteinQ1} we have
\begin{equation}\label{error2V}
\begin{split}
\beta(\alpha,{\mathbf{V}}_{n}) = {\mathbf Q}_{{\mathbf V}_{n}}\left(
{\cal A}_{\mu_{0}}\right) = \int\limits_{{\cal A}_{\mu_{0}}}
p_{{\mathbf V}_{n}}({\mathbf x})d{\mathbf x}  
= \int\limits_{{\cal A}_{\mu_{0}}}\frac{p_{{\mathbf V}_{n}}}
{p_{{\mathbf M}_{n}}}({\mathbf x})
\frac{p_{{\mathbf M}_{n}}}{p_{{\mathbf I}_{n}}}
({\mathbf x})p_{{\mathbf I}_{n}}({\mathbf x})d{\mathbf x} = \\
= e^{-D({\mathbf I}_{n}||{\mathbf M}_{n})+\mu_{2}}
\int\limits_{{\cal A}_{\mu_{0}}}\frac{p_{{\mathbf V}_{n}}}{p_{{\mathbf M}_{n}}}
({\mathbf x})p_{{\mathbf I}_{n}}({\mathbf x})d{\mathbf x}  
\leq \beta(\alpha,{\mathbf{M}}_{n})e^{\mu_{2}-\mu_{1}}
{\mathbf E}_{{\mathbf I}_{n}}\frac{p_{{\mathbf V}_{n}}}{p_{{\mathbf M}_{n}}}
({\mathbf x}),
\end{split}
\end{equation}
where $0 \leq \max\{\mu_{1},\mu_{2}\} \leq \mu_{0}(\alpha,\mathbf M_{n})$.
By assumption \eqref{assump3} and \eqref{lem1a} we have
\begin{equation}\label{error2V1a}
\begin{gathered}
\mu_{0}(\alpha,\mathbf M_{n}) = O(n^{1/(1+\delta)})  = \varepsilon(n),
\quad n\to \infty.
\end{gathered}
\end{equation}
Therefore, if
\begin{equation}\label{error2V1}
\begin{gathered}
\sup_{{\mathbf V}_{n} \in \mathcal{M}_{n}^{LR}(\mathbf{M}_{n})}
{\mathbf E}_{{\mathbf I}_{n}}\frac{p_{{\mathbf V}_{n}}}{p_{{\mathbf M}_{n}}}
({\mathbf x}) \leq e^{\varepsilon(n)}, \quad n\to \infty,
\end{gathered}
\end{equation}
then by \eqref{error2V}-\eqref{error2V1}, as $n\to \infty$
\begin{equation}\label{error2V2}
\begin{gathered}
\sup_{{\mathbf V}_{n} \in \mathcal{M}_{n}^{LR}(\mathbf{M}_{n})}
\ln\beta(\alpha,{\mathbf{V}}_{n}) \leq
\ln\beta(\alpha,{\mathbf{M}}_{n}) + \varepsilon(n).
\end{gathered}
\end{equation}

\subsection{``Outer'' bound}
Now, we get a similar ``outer bound'' for $\mathcal{M}_{n}(\mathbf{M}_{n})$.
Let ${\cal D} \in \mathbf{R}^{n}$ be a decision region in favor of
$\mathbf{I}_{n}$, and $\beta_{\mathbf{M}_{n}}=\beta_{\mathbf{M}_{n}}({\cal D})$,
$\alpha = \alpha({\cal D})$ be the corresponding error probabilities.
Then denoting $p=p_{\mathbf{I}_{n}}$ and  $q=p_{\mathbf{M}_{n}}$,
similarly to \eqref{Defab}, we have
\begin{equation}\label{Defab1}
\begin{split}
\beta_{\mathbf{M}_{n}} = {\mathbf Q}_{\mathbf{M}_{n}}({\cal D}) =
\int\limits_{{\cal D}}q({\mathbf x})d{\mathbf x}, \qquad
\alpha = {\mathbf P}_{\mathbf{I}_{n}}({\cal D}^{c}).
\end{split}
\end{equation}
Consider an another matrix $\mathbf{V}_{n}\in \mathcal M_{n}(\mathbf{M}_{n})$.
Denoting $q_{1}=p_{\mathbf{V}_{n}}$, we must have for the 2-nd error
probability $\beta_{\mathbf{V}_{n}} = \beta_{\mathbf{V}_{n}}({\cal D})$
\begin{equation}\label{Defab1a}
\begin{gathered}
\beta_{\mathbf{V}_{n}} = {\mathbf Q}_{\mathbf{V}_{n}}({\cal D}) =
\int\limits_{{\cal D}}q_{1}({\mathbf x})d{\mathbf x} \leq
\beta_{\mathbf{M}_{n}} e^{\varepsilon(n)}.
\end{gathered}
\end{equation}
For some $\delta$, $0 \leq \delta \leq 1$, consider also the probability density
function $q_{\delta}({\mathbf x})$
\begin{equation}\label{Defab2}
\begin{gathered}
q_{\delta}({\mathbf x}) = (1-\delta)q({\mathbf x}) + \delta q_{1}({\mathbf x}),
\end{gathered}
\end{equation}
and the following value $\beta_{\delta}$ for it
\begin{equation}\label{Defab3a}
\begin{gathered}
\beta_{\delta} = \int\limits_{{\cal D}}q_{\delta}({\mathbf x})d{\mathbf x} =
(1-\delta)\beta_{\mathbf{M}_{n}} + \delta\beta_{\mathbf{V}_{n}}.
\end{gathered}
\end{equation}
By \eqref{Defab1}-\eqref{Defab1a} we have
\begin{equation}\label{Defab3}
\begin{gathered}
\beta_{\delta} \leq \beta_{\mathbf{V}_{n}}
\left(1-\delta + \delta e^{\varepsilon(n)}\right).
\end{gathered}
\end{equation}
It may be noted that the probability density $q_{\delta}({\mathbf x})$
corresponds to the Bayes problem statement, when the alternative hypothesis
$\mathcal H_{1}$ with probability $(1-\delta)$ coincides with $\mathbf{M}_{n}$,
and with probability $\delta$ is $\mathbf{V}_{n}$. Respectively, the value
$\beta_{\delta}$ is the 2-nd kind error probability.

Similarly to \eqref{Stein1a}-\eqref{Stein1c}, we lowerbound the value
$\beta_{\delta}$. First, similarly to \eqref{Stein1a}, we have
\begin{equation}\label{Defab4}
\begin{gathered}
\ln\frac{\beta_{\delta}}{1-\alpha} = \ln\left[\frac{1}{(1-\alpha)}
\int\limits_{{\cal D}}p({\mathbf x})\frac{q_{\delta}}{p}({\mathbf x})d{\mathbf x}
\right] \geq \frac{1}{(1-\alpha)}
\int\limits_{{\cal D}}p({\mathbf x})\ln\frac{q_{\delta}}{p}({\mathbf x})
d{\mathbf x} = \\
= - \frac{D(p({\mathbf x})||q_{\delta}({\mathbf x}))}{1-\alpha} -
\frac{1}{(1-\alpha)}\int\limits_{{\cal D}^{c}}p({\mathbf x})
\ln\frac{q_{\delta}}{p}({\mathbf x})d{\mathbf x}.
\end{gathered}
\end{equation}
Similarly to \eqref{Stein1b}, for the last term in the right-hand side of
\eqref{Defab4} we have
\begin{equation}\label{Defab5}
\begin{split}
\int\limits_{{\cal D}^{c}}p({\mathbf x})
\ln\frac{q_{\delta}}{p}({\mathbf x})d{\mathbf x} \leq
\alpha\ln\left[\frac{1}{\alpha}
\int\limits_{{\cal D}^{c}}q_{\delta}({\mathbf x})d{\mathbf x}\right] 
= \alpha\ln\frac{1-\beta_{\delta}}{\alpha}
\leq \alpha\ln\frac{1}{\alpha}.
\end{split}
\end{equation}
Therefore, similarly to \eqref{Stein11}, we get
\begin{equation}\label{Stein11a}
\begin{gathered}
\ln\beta_{\delta} \geq
-\frac{D(p({\mathbf x})||q_{\delta}({\mathbf x}))+h(\alpha)}{1-\alpha}.
\end{gathered}
\end{equation}

Consider the value $D(p({\mathbf x})||q_{\delta}({\mathbf x}))$ from
the right-hand side of \eqref{Stein11a}. Denoting
\begin{equation}\label{Defr}
\begin{gathered}
r({\mathbf x}) = \frac{q_{1}({\mathbf x})}{q({\mathbf x})},
\end{gathered}
\end{equation}
by \eqref{Defab2} we have
\begin{equation}\label{Defab6}
\begin{gathered}
\frac{q_{\delta}({\mathbf x})}{q({\mathbf x})} =
1-\delta + \delta\frac{q_{1}({\mathbf x})}{q({\mathbf x})} =
1-\delta + \delta r({\mathbf x}).
\end{gathered}
\end{equation}
Therefore
\begin{equation}\label{Defab7}
\begin{split}
D(p({\mathbf x})||q_{\delta}({\mathbf x})) =
-\int\limits_{\mathbf{R}^{n}} p({\mathbf x})
\ln\frac{q_{\delta}}{p}({\mathbf x})d{\mathbf x}  
= D(p({\mathbf x})||q({\mathbf x})) + g(\delta),
\end{split}
\end{equation}
where
\begin{equation}\label{Defab8}
\begin{gathered}
g(\delta) = - \int\limits_{\mathbf{R}^{n}} p(\mathbf x)\ln\left[1-\delta +
\delta r(\mathbf x)\right]d\mathbf x.
\end{gathered}
\end{equation}
Then, by \eqref{Defab3} and \eqref{Defab7}-\eqref{Defab8}, we must have 
\begin{equation}\label{Defab9}
\begin{gathered}
g(\delta) \geq -\ln\left(1-\delta+\delta e^{\varepsilon(n)}\right), \quad
\text{for all } \ 0 < \delta \leq 1.
\end{gathered}
\end{equation}
Note that by inequality $\ln {\mathbf E}\xi \geq {\mathbf E}\ln \xi$,
we have from \eqref{Defab8}
\begin{equation}\label{Defab91}
\begin{gathered}
g(\delta) \leq \ln\int\limits_{\mathbf{R}^{n}}
\frac{p(\mathbf x)}{1-\delta +\delta r(\mathbf x)}d\mathbf x, \quad
\text{for all } \ 0 < \delta \leq 1.
\end{gathered}
\end{equation}
Therefore, in order to have \eqref{Defab9} fulfilled, we need to have
\begin{equation}\label{Defab92}
\begin{gathered}
\int\limits_{\mathbf{R}^{n}}
\frac{p(\mathbf x)}{1-\delta +\delta r(\mathbf x)}d\mathbf x \geq
\frac{1}{1-\delta+\delta e^{o(n)}}, \quad 0 < \delta \leq 1.
\end{gathered}
\end{equation}
Since $\int p(\mathbf x)d\mathbf x=1$, relation \eqref{Defab92} is equivalent to
\begin{equation}\label{Defab92a}
\begin{gathered}
\int\limits_{\mathbf{R}^{n}}
\frac{p(\mathbf x)(r(\mathbf x)-1)}{1-\delta +\delta r(\mathbf x)}d\mathbf x
\leq \frac{e^{o(n)}-1}{1-\delta+\delta e^{o(n)}}, \quad 0 < \delta \leq 1.
\end{gathered}
\end{equation}
Note that
\begin{equation}\label{Defab93}
\begin{gathered}
\int\limits_{\mathbf{R}^{n}}
\frac{p(\mathbf x)}{1-\delta +\delta r(\mathbf x)}d\mathbf x \leq
\frac{1}{1-\delta}, \quad 0 < \delta \leq 1.
\end{gathered}
\end{equation}
Then, in order to have fulfilled \eqref{Defab92a}, we need, at least,
\begin{equation}\label{Defab94}
\begin{gathered}
\int\limits_{\mathbf{R}^{n}}
\frac{p(\mathbf x)r(\mathbf x)}{1 +\delta r(\mathbf x)}d\mathbf x
\leq \frac{e^{o(n)}}{(1-\delta)(1-\delta+\delta e^{o(n)})},
\quad \text{for all } \ 0 < \delta \leq 1.
\end{gathered}
\end{equation}
Setting $\delta \downarrow 0$, we get from \eqref{Defab94} the necessary 
condition 
\begin{equation}\label{Defab98}
\begin{gathered}
\int\limits_{\mathbf{R}^{n}}p(\mathbf x)r(\mathbf x)d\mathbf x =
{\mathbf E}_{{\mathbf I}_{n}}\frac{p_{{\mathbf V}_{n}}}{p_{{\mathbf M}_{n}}}
({\mathbf x}) \leq e^{o(n)},
\end{gathered}
\end{equation}
then coincides with \eqref{error2V1} for the matrix ${\mathbf M}_{n}$.
In the case of the set $\mathcal{M}_{n}(\mathbf{M}_{n})$ the condition
\eqref{Defab98} should be fulfilled for all 
$\mathbf{V}_{n} \in \mathcal{M}_{n}(\mathbf{M}_{n})$, i.e. it is necessary 
to have
\begin{equation}\label{condition5}
\begin{split}
\sup\limits_{\mathbf{V}_{n} \in \mathcal{M}_{n}(\mathbf{M}_{n})}
{\mathbf E}_{\mathbf{I}_{n}}\frac{p_{\mathbf{V}_{n}}}
{p_{\mathbf{M}_{n}}}({\mathbf x}) \leq e^{o(n)},
\end{split}
\end{equation}
from which the ``outer bound'' for
$\mathcal{M}_{n}(\mathbf{M}_{n})$ follows (see \eqref{Theor1}).

It remains us to express analytically the condition in \eqref{Defab98} via 
matrices $\mathbf{M}_{n},\mathbf{V_{n}}$ (see \eqref{Theor1a}-\eqref{Theor1}). 
By \eqref{deGaus2} we have
\begin{equation}\label{Stein3b41k}
\begin{split}
& \mathbf E_{\boldsymbol{\xi}_{n}}\frac{p_{\mathbf V_{n}}}
{p_{\mathbf M_{n}}}\left(\boldsymbol{\xi}_{n}\right) = 
\frac{\left|\mathbf{M}_{n}\right|^{1/2}}
{\left|\mathbf{V}_{n}\right|^{1/2}}
{\mathbf E}_{\boldsymbol{\xi}_{n}}
e^{-\left(\boldsymbol{\xi}_{n},\left(\mathbf{V}_{n}^{-1} -
\mathbf{M}_{n}^{-1}\right)\boldsymbol{\xi}_{n}\right)/2}.
\end{split}
\end{equation}

Note that if a matrix $\mathbf{I}_{n}+\mathbf{A}_{n}$ is positive definite,
then with
$\boldsymbol{\xi}_{n} \sim {\mathcal N}({\mathbf 0}_{n},{\mathbf I}_{n})$,
from \cite[Ch. 6.9, Theorem 3]{Bellman} we have
\begin{equation}\label{Stein3b31h}
\begin{gathered}
{\mathbf E}_{\boldsymbol{\xi}_{n}}
e^{-(\boldsymbol{\xi}_{n},\mathbf{A}_{n}\boldsymbol{\xi}_{n})/2} =
\frac{1}{\left|\mathbf{I}_{n}+\mathbf{A}_{n}\right|^{1/2}}.
\end{gathered}
\end{equation}
If a matrix $\mathbf{I}_{n}+\mathbf{A}_{n}$ is not positive definite, then
\begin{equation}\label{Stein3b31t}
\begin{gathered}
{\mathbf E}_{\boldsymbol{\xi}_{n}}
e^{-(\boldsymbol{\xi}_{n},\mathbf{A}_{n}\boldsymbol{\xi}_{n})/2} = \infty.
\end{gathered}
\end{equation}

Assume first a matrix $\mathbf{I}_{n}+\mathbf{V}_{n}^{-1}-\mathbf{M}_{n}^{-1}$
be positive definite. Then, by \eqref{Stein3b41k} and \eqref{Stein3b31h}
\begin{equation}\label{Stein3b31k}
\begin{split}
&\mathbf E_{\boldsymbol{\xi}_{n}}\frac{p_{\mathbf V_{n}}}
{p_{\mathbf M_{n}}}\left(\boldsymbol{\xi}_{n}\right) =
\frac{\left|\mathbf{M}_{n}\right|^{1/2}}
{\left|\mathbf{V}_{n}\right|^{1/2}\left|\mathbf{I}_{n}+\mathbf{V}_{n}^{-1} -
\mathbf{M}_{n}^{-1}\right|^{1/2}} =
\frac{\left|\mathbf{M}_{n}\right|^{1/2}}{\left|\mathbf{I}_{n}+
\mathbf{V}_{n}\left(\mathbf{I}_{n}-\mathbf{M}_{n}^{-1}\right)\right|^{1/2}}.
\end{split}
\end{equation}
Therefore, by \eqref{Stein3b31k}, condition  \eqref{condition5}
is equivalent to the relation
\begin{equation}\label{Stein3b31n}
\begin{gathered}
\sup\limits_{\mathbf{V}_{n} \in \mathcal{M}_{n}(\mathbf{M}_{n})}
\frac{\left|\mathbf{M}_{n}\right|}{\left|\mathbf{I}_{n}+
\mathbf{V}_{n}\left(\mathbf{I}_{n}-\mathbf{M}_{n}^{-1}\right)\right|}
\leq e^{o(n)},
\end{gathered}
\end{equation}
provided a matrix $\mathbf{I}_{n}+\mathbf{V}_{n}^{-1} -\mathbf{M}_{n}^{-1}$ is
positive definite.

If a matrix $\mathbf{I}_{n}+\mathbf{V}_{n}^{-1} -\mathbf{M}_{n}^{-1}$ is not
positive definite, then by \eqref{Stein3b41k} and \eqref{Stein3b31t}
\begin{equation}\label{Stein3b31m}
\begin{split}
&\mathbf E_{\boldsymbol{\xi}_{n}}\frac{p_{\mathbf V_{n}}}
{p_{\mathbf M_{n}}}\left(\boldsymbol{\xi}_{n}\right) = \infty,
\end{split}
\end{equation}
and therefore condition \eqref{condition5} is not satisfied.

We define $\mathcal{M}_{n}(\mathbf{M}_{n})$ as the largest set
satisfying condition \eqref{Stein3b31n}.
The set $\mathcal{M}_{n}(\mathbf{M}_{n})$ coincides with
definition \eqref{Theor1a}-\eqref{Theor1}.
From \eqref{error2V2}, \eqref{condition5} and \eqref{Stein3b31k},
Theorem 1 follows. $\qquad \Box$

\subsection{Proof of Theorem 2}
We develop the left-hand side of relation \eqref{Stein3b31n} as follows.
For a covariance matrix $\mathbf{M}_{n}$ with eigenvalues $\{\lambda_{i}\}$,
let us consider covariance matrices $\mathbf{V}_{n}$, commutating with
$\mathbf{M}_{n}$, i.e.
$\mathbf{M}_{n}\mathbf{V}_{n}=\mathbf{V}_{n}\mathbf{M}_{n}$. Then each pair
$\mathbf{M}_{n},\mathbf{V}_{n}$ has the same set of eigenvectors
$\{\mathbf{x}_{i}\}$ \cite[Ch. 4.11, Theorem 5]{Bellman}. Denote by
$\{\nu_{i}\}$ the eigenvalues of $\mathbf{V}_{n}$.
Then the matrix $\mathbf{B}_{n} = \mathbf{I}_{n}+
\mathbf{V}_{n}\left(\mathbf{I}_{n}-\mathbf{M}_{n}^{-1}\right)$
has eigenvalues
\begin{equation}\label{Theor1d}
\begin{split}
1+\nu_{i}-\nu_{i}/\lambda_{i}, \qquad i=1,\ldots,n.
\end{split}
\end{equation}
Therefore,
\begin{equation}\label{Theor1d2}
\begin{split}
f(\mathbf{M}_{n},{\mathbf V}_{n}) =
\prod_{i=1}^{n}\frac{[\lambda_{i}+\nu_{i}(\lambda_{i}-1)}{\lambda_{i}^{2}},
\end{split}
\end{equation}
from where Theorem 2 follows. $\qquad \Box$

\subsection{Proof of Corollary 1}
Let $\boldsymbol{\xi}_{n}$ be a
Gaussian random vector with $\boldsymbol{\xi}_{n} \sim
{\mathcal N}(\boldsymbol{0},\mathbf{I}_{n})$ and $\mathbf{A}$ be a
symmetric $(n \times n)$-matrix with eigenvalues $\{a_{i}\}$.
Consider the quadratic form
$(\boldsymbol{\xi}_{n},\mathbf{A}\boldsymbol{\xi}_{n})$. There
exists the orthogonal $(n \times n)$-matrix $\mathbf{T}$, such that
$\mathbf{T}'\mathbf{A}\mathbf{T} = \mathbf{B}$, where
$\mathbf{B}$ is the diagonal matrix with diagonal elements
$\{a_{i}\}$ \cite[Ch. 4.7]{Bellman}. Since
$\mathbf{T}\boldsymbol{\xi}_{n} \sim
{\mathcal N}(\boldsymbol{0},\mathbf{I}_{n})$, the quadratic forms
$(\boldsymbol{\xi}_{n},\mathbf{A}\boldsymbol{\xi}_{n})$ and
$(\boldsymbol{\xi}_{n},\mathbf{B}\boldsymbol{\xi}_{n})$ have the
same distributions. Therefore, by formulas \eqref{deGaus3} and
\eqref{Stein2} we have
\begin{equation}\label{Stein1bb}
\begin{gathered}
\ln\frac{p_{\mathbf{I}_{n}}}{p_{\mathbf{M}_{n}}}
\left({\mathbf y}_{n}\right)\stackrel{d}{=}
\frac{1}{2}\left[\ln|\mathbf{M}_{n}| + \zeta_{n}\right],
\end{gathered}
\end{equation}
where
\begin{equation}\label{Stein1bc}
\begin{gathered}
\zeta_{n}= \left({\mathbf y}_{n},
\left[\mathbf{M}_{n}^{-1} -\mathbf{I}_{n}\right]{\mathbf y}_{n}\right),
\end{gathered}
\end{equation}
and
\begin{equation*}
\begin{split}
D(\mathbf{I}_{n}||\mathbf{M}_{n}) =
\frac{1}{2}\sum_{i=1}^{n}\left(\ln\lambda_{i} +\frac{1}{\lambda_{i}}-1\right),
\end{split}
\end{equation*}
where $\{\lambda_{i}\}$ are the eigenvalues of the matrix $\mathbf{M}_{n}$
(the eigenvalues of the matrix $\mathbf{M}_{n}^{-1}$ are $\{\lambda_{i}^{-1}\}$).
We use the following result \cite[Ch. III.5.15]{Petrov}: let
$\zeta_{1},\ldots,\zeta_{n}$ be independent random variables with
${\mathbf{E}}\zeta_{i}=0$, $i=1,\ldots,n$. Then for any $1\leq p\leq2,$
\begin{equation}\label{BarEs1}
\begin{gathered}
{\mathbf E}\left|\sum_{i=1}^{n}\zeta_{i}\right| \leq
2\sum_{i=1}^{n}{\mathbf E}|\zeta_{i}|^{p},
\end{gathered}
\end{equation}
provided the right-hand side of \eqref{BarEs1} is finite.

For any $1\leq p\leq 2,$ we use Chebychev inequality  and
\eqref{Stein1bb}--\eqref{BarEs1}, which give (see \eqref{SteinP1a})
\begin{equation}\label{lem1}
\begin{split}
&\alpha_{\mu} = {\mathbf P}_{\mathbf{I}_{n}}\left\{
\ln\frac{p_{\mathbf{I}_{n}}}{p_{\mathbf{M}_{n}}}({\mathbf x}) \leq
D({\mathbf I}_{n}||{\mathbf M}_{n})-\mu\right\}  \\
&\leq {\mathbf P}_{\mathbf{I}_{n}}\left\{\left|
\ln\frac{p_{\mathbf{I}_{n}}}{p_{\mathbf{M}_{n}}}
({\mathbf x}) - D({\mathbf I}_{n}||{\mathbf M}_{n})\right|> \mu\right\}  \\
&= {\mathbf P}_{\boldsymbol{\xi}_{n}}\left\{\left|\zeta_{n} -
\sum_{i=1}^{n}\left(\frac{1}{\lambda_{i}}-1\right)\right| > 2\mu\right\}  \\
&= {\mathbf P}_{\boldsymbol{\xi}_{n}}\left\{\left|
\sum_{i=1}^{n}\left(\frac{1}{\lambda_{i}}-1\right)(\xi_{i}^{2}-1)
\right| > 2\mu\right\} \\
&\leq 2(2\mu)^{-p}\sum_{i=1}^{n}
{\mathbf E}\left|\left(\frac{1}{\lambda_{i}}-1\right)(\xi_{i}^{2}-1)
\right|^{p} \\
&\leq 2^{1-p}3^{p/2}\mu^{-p}\sum_{i=1}^{n}
\left|\frac{1}{\lambda_{i}}-1\right|^{p} \leq 3C_{p}\mu^{-p}n.
\end{split}
\end{equation}
Then, from condition $\alpha_{\mu} \leq \alpha$ and \eqref{lem1}
we get formula \eqref{lem1a}.  $\qquad\Box$

\section{Model \eqref{mod2} and  Stationary Stochastic Signals}

\subsection{Model \eqref{mod2}}

Since model \eqref{mod2} is a particular case of model \eqref{mod1}, Theorem 1
and Theorem 2 can also be applied to this model.  Let $\mathbf{s}_{n}$ be a
``stochastic signal'' independent on $\boldsymbol{\xi}_{n}$ and having the
distribution $\mathbf{s}_{n}\sim
{\mathcal{N}}(\boldsymbol{0},\mathbf{S}_{n})$. Let also $\mathcal{S}_{n}$ be a
given set of covariance matrices $\mathbf{S}_{n}$. Then $\mathbf{M}%
_{n}=\mathbf{S}_{n}+\mathbf{I}_{n}$. Denote by $\{\mu_{i}(\mathbf{S}_{n})\})$
the eigenvalues (all positive) of the covariance matrix $\mathbf{S}_{n}$.
Then $\mu_{i}(\mathbf{S}_{n})=\lambda_{i}(\mathbf{M}_{n})-1$, $i=1,\ldots,n$.
Instead of assumptions \eqref{assump0}-\eqref{assump3},  we use their analogs:

{\bf III}. For all covariance matrices
$\mathbf{S}_{n} \in {\cal S}_{n}$ there exist limits
\begin{equation}\label{assump0a}
\begin{gathered}
\lim_{n\to\infty}\frac{1}{n}\sum_{i=1}^{n}
\left[\ln(\mu_{i}(\mathbf{S}_{n})+1) +
\frac{1}{\mu_{i}(\mathbf{S}_{n})+1}-1\right],
\end{gathered}
\end{equation}
where convergence is uniform on $\mathbf{S}_{n} \in {\cal S}_{n}$.

{\bf IV}. For some $\delta > 0$ we have
\begin{equation}\label{assump3a}
\begin{gathered}
\lim_{n\to\infty}\frac{1}{n}\sup_{\mathbf{S}_{n} \in {\cal S}_{n}}
\sum_{i=1}^{n}\left[
\frac{\mu_{i}(\mathbf{S}_{n})}{\mu_{i}(\mathbf{S}_{n})+1}
\right]^{1+\delta} < \infty.
\end{gathered}
\end{equation}

Instead of the function $f(\mathbf{M}_{n},{\mathbf V}_{n})$ from \eqref{Theor1a},
we introduce its analog $t(\mathbf{S}_{n},{\mathbf V}_{n})$. For any
$\mathbf{S}_{n},{\mathbf{V}}_{n}\in\mathcal{C}_{n},$ such that
$\mathbf{I}_{n}+(\mathbf{I}_{n}+\mathbf{V}_{n})^{-1} -
(\mathbf{I}_{n}+\mathbf{S}_{n})^{-1} > 0$, define the function
\begin{equation}\label{Theor121}
\begin{split}
t(\mathbf{S}_{n},{\mathbf V}_{n}) =
\frac{\left|\mathbf{I}_{n}+\mathbf{S}_{n} +
(\mathbf{V}_{n}-\mathbf{S}_{n})
\mathbf{S}_{n}(\mathbf{I}_{n}+\mathbf{S}_{n})^{-1}\right|}
{\left|\mathbf{I}_{n}+\mathbf{S}_{n}\right|}.
\end{split}
\end{equation}
In derivation of \eqref{Theor121} we used a simple formula
$$
\mathbf{I}_{n}-(\mathbf{I}_{n}+\mathbf{S}_{n})^{-1} =
\mathbf{S}_{n}(\mathbf{I}_{n}+\mathbf{S}_{n})^{-1}.
$$

As a direct consequence of Theorem 1, we get

{\bf Corollary 2}. \textit{
If assumptions \eqref{assump0a} and
\eqref{assump3a} are satisfied for model \eqref{mod2}, then the largest set
$\mathcal{S}_{n}(\mathbf{S}_{n})$ that satisfies asymptotic equality
\begin{equation}\label{noise1}
\begin{gathered}
\lim_{n\to\infty}\frac{1}{n}\ln\beta(\mathbf{S}_{n}) =
\lim_{n\to\infty}\frac{1}{n}\ln\beta({\cal S}_{n}),
\end{gathered}
\end{equation}
for $n\rightarrow\infty$ has the form}
\begin{equation}\label{Theor101}
\begin{split}
&{\cal S}_{n}(\mathbf{S}_{n}) = \left\{\mathbf{V}_{n}:
\begin{array}{c}
\mathbf{I}_{n}+(\mathbf{I}_{n}+\mathbf{V}_{n})^{-1} -
(\mathbf{I}_{n}+\mathbf{S}_{n})^{-1} > 0, \\
\sup\limits_{\mathbf{V}_{n} \in {\cal S}_{n}(\mathbf{S}_{n})}
t(\mathbf{S}_{n},{\mathbf V}_{n}) \leq e^{o(n)}
\end{array}
\right\}.
\end{split}
\end{equation}

Clearly, the set ${\cal S}_{n}(\mathbf{S}_{n})$ is convex on
${\mathbf V}_{n}$.

We simplify Corollary 2 using Theorem 2 as follows. For the matrix
$\mathbf{S}_{n}$ with the eigenvalues $\{\mu_{i}\}$, consider in
\eqref{Theor101} only those  matrices $\mathbf{V}_{n}$, that
commutate with $\mathbf{S}_{n}$. Denote by $\{\nu_{i}\}$ the
eigenvalues of $\mathbf{V}_{n}$. Then, similarly to \eqref{deffVM}, we get
\begin{equation}\label{Theor122}
\begin{split}
t(\mathbf{S}_{n},{\mathbf V}_{n}) = \prod_{i=1}^{n}\left[1+
\frac{(\nu_{i}-\mu_{i})\mu_{i}}{(1+\mu_{i})^{2}}\right].
\end{split}
\end{equation}

Similarly to \eqref{defC}, for a matrix $\mathbf{S}_{n}$ introduce the convex set
${\cal C}_{\mathbf{S}_{n}}$ of covariance matrices $\mathbf{V}_{n}$,
commutating with $\mathbf{S}_{n}$:
\begin{equation}\label{defCS}
\begin{split}
{\cal C}_{\mathbf{S}_{n}} =\left\{\mathbf{V}_{n}:
\mathbf{S}_{n}\mathbf{V}_{n} =\mathbf{V}_{n}\mathbf{S}_{n}\right\}.
\end{split}
\end{equation}
We also introduce the following subset of ${\cal C}_{\mathbf{S}_{n}}$
(see \eqref{Theor122})
\begin{equation}\label{defV2}
\begin{split}
{\cal V}_{n}^{(2)}(\mathbf{S}_{n}) = 
\Bigg\{{\mathbf V}_{n} \in {\cal C}_{\mathbf{S}_{n}}:
\sup_{\mathbf{V}_{n} \in {\cal V}_{n}^{(2)}(\mathbf{S}_{n})}
t(\mathbf{S}_{n},{\mathbf V}_{n})\leq e^{o(n)} \Bigg\}.
\end{split}
\end{equation}
The set $\mathcal{V}_{n}^{(2)}(\mathbf{S}_{n})$ is convex on $\mathbf{V}_{n}$,
since the function $\ln z$ is concave on $z>0$. Then, similarly to Theorem 2,
we get the following ``inner'' bound $\mathcal{V}_{n}^{(2)}(\mathbf{S}_{n})$
for $\mathcal{S}_{n}(\mathbf{S}_{n})$:

{\bf Corollary 3}. \textit{Let assumptions \eqref{assump0a}-\eqref{assump3a}
be satisfied for model \eqref{mod2}. Then  for
the largest set $\mathcal{S}_{n}(\mathbf{S}_{n})$ such that
formula \eqref{noise1} holds, we have
\begin{equation}\label{Cor3}
\begin{gathered}
{\cal V}_{n}^{(2)}(\mathbf{S}_{n})\subseteq
{\cal S}_{n}(\mathbf{S}_{n}),
\end{gathered}
\end{equation}
where the set ${\cal V}_{n}^{(2)}(\mathbf{S}_{n})$ is
defined in \eqref{defV2}.}

\subsection{Stationary Stochastic Signals}

Consider model \eqref{mod2}, where $\mathbf{s}_{n}\sim{\mathcal{N}}
(\boldsymbol{0},\mathbf{S}_{n})$ is a wide-sense stationary Gaussian
process with the mean zero and the power spectral density
$f_{\mathbf{S}_{n}}(\omega),\omega\in\lbrack\pi,\pi]$.
Let $\mathcal{K}_{n}$ be a given set of covariance matrices
$\mathbf{K}_{n}$ that are competitive to $\mathbf{S}_{n}$.

We consider only  the case when covariance matrices $\mathbf{K}_{n}$
with power spectral densities $f_{\mathbf{K}_{n}}(\omega)$ commutate
with $\mathbf{S}_{n}$. Then we are able to apply Corollary 3, which
can be expressed via power spectral densities.
By Theorem of Szeg\"{o} \cite[Ch. 5.2, Theorem]{GS},
\cite[Theorem 4.1]{Gray}, we replace assumptions \eqref{assump0a}
and \eqref{assump3a} by their analogs:

{\bf V}. For all power spectral densities $f_{\mathbf{K}_{n}}(\omega)$,
$\mathbf{K}_{n} \in {\cal K}_{n}$, there exist finite limits
\begin{equation}\label{assump4}
\begin{gathered}
\lim_{n\to \infty}
\int\limits_{-\pi}^{\pi}\left[\ln(f_{\mathbf{K}_{n}}(\omega)+1) +
\frac{1}{f_{\mathbf{K}_{n}}(\omega)+1} -1\right]d\omega,
\end{gathered}
\end{equation}
where convergence is uniform on $\mathbf{K}_{n} \in {\cal K}_{n}$.

{\bf VI}. For all power spectral densities
$f_{\mathbf{K}_{n}}(\omega)$ and some $\delta > 0$
\begin{equation}\label{assump4a}
\begin{gathered}
\lim_{n\to \infty}\sup_{\mathbf{K}_{n}\in {\cal K}_{n}}
\int\limits_{-\pi}^{\pi}\left|\frac{f_{\mathbf{K}_{n}}(\omega)}
{f_{\mathbf{K}_{n}}(\omega)+1}\right|^{1+\delta} < \infty.
\end{gathered}
\end{equation}

For a given power spectral density $f_{\mathbf{S}_{n}}(\omega),$  denote by
$\mathcal{F}_{n}(f_{\mathbf{S}_{n}})$ the largest set of power spectral
densities,  which satisfy the following analog of equality \eqref{aseq1}
\begin{equation}\label{stat2}
\lim_{n\to\infty}\frac{1}{n}\ln\beta(f_{\mathbf{S}_{n}}) =
\lim_{n\to\infty}\frac{1}{n}\ln\beta(
{\cal F}_{n}(f_{\mathbf{S}_{n}})).
\end{equation}
In other words, for a given 1-st kind error probability $\alpha$,
$\mathcal{F}_{n}(f_{\mathbf{S}_{n}})$ is the maximal set of power
spectral densities, which can be replaced by the density
$f_{\mathbf{S}_{n}}$ (without asymptotic loss for
$\beta(\mathcal{F}_{n}(f_{\mathbf{S}_{n}}))$).

In order to describe an ``inner bound'' for the set
${\cal F}_{n}(f_{\mathbf{S}_{n}})$, introduce the following
functional:
\begin{equation}\label{stoch3b}
\begin{split}
f({\mathbf{S}_{n}},{\mathbf{K}_{n}}) = 
\int\limits_{-\pi}^{\pi}\ln\left[1+\frac{f_{\mathbf{S}_{n}}(\omega)
[f_{\mathbf{K}_{n}}(\omega)-f_{\mathbf{S}_{n}}(\omega)]}
{(1+f_{\mathbf{S}_{n}}(\omega))^{2}}\right]d\omega.
\end{split}
\end{equation}
The functional $f({\mathbf{S}_{n}},{\mathbf{K}_{n}})$ is the analog of
the functional $\ln t(\mathbf{S}_{n},{\mathbf V}_{n})$ from
\eqref{Theor122}, which follows by Theorem of Szeg\"{o}.

Introduce also the set (as $n\to \infty$)
\begin{equation}\label{stat1}
\begin{gathered}
{\cal F}_{n}^{(1)}(f_{\mathbf{S}_{n}}) = \left\{
f_{\mathbf{K}_{n}}(\omega):
f({\mathbf{S}_{n}},{\mathbf{K}_{n}}) \leq o(1)\right\},
\end{gathered}
\end{equation}
where the inequality in the right-hand side of \eqref{stat1}
fulfills uniformly over $f_{\mathbf{K}_{n}}(\omega) \in
{\cal F}_{n}^{(1)}(f_{\mathbf{S}_{n}})$.
The set ${\cal F}_{n}^{(1)}(f_{\mathbf{S}_{n}})$ is the analog of
the set ${\cal V}_{n}^{(2)}(\mathbf{S}_{n})$ from \eqref{defV2}.

As a direct consequence of Corollary 3, we get

{\bf Corollary 4}. \textit{If assumptions \eqref{assump4}--\eqref{assump4a}
are satisfied for model \eqref{mod2},  then for the
largest set $\mathcal{F}_{n}(f_{\mathbf{S}_{n}})$ for which formula
\eqref{stat2} holds, we have
\begin{equation}\label{Cor4}
\begin{gathered}
{\cal F}_{n}^{(1)}(\mathbf{S}_{n}) \subseteq
{\cal F}_{n}(f_{\mathbf{S}_{n}}),
\end{gathered}
\end{equation}
where the set of densities ${\cal F}_{n}^{(1)}(\mathbf{S}_{n})$ is
defined in \eqref{stat1}.}

The set ${\cal F}_{n}^{(1)}(\mathbf{S}_{n})$ is convex in
$f_{\mathbf{K}_{n}}(\omega)$, since the \\
function $\ln z$ is concave
in $z > 0$. In other words, if
$f_{\mathbf{K}_{n}^{(0)}}(\omega), f_{\mathbf{K}_{n}^{(1)}}(\omega)
\in {\cal F}_{n}^{(1)}(\mathbf{S}_{n})$, then
$cf_{\mathbf{K}_{n}^{(0)}}(\omega)+
(1-c)f_{\mathbf{K}_{n}^{(1)}}(\omega)
\in {\cal F}_{n}^{(1)}(\mathbf{S}_{n})$, for any $0 \leq c \leq 1$.

{\it Remark 4}. In \cite[Theorem 1]{ZhangPoor11} similar to
\eqref{stat1}, condition was derived (with $0$ instead of $o(1)$).

\section{Examples}
{\bf Example 1}. In some symmetric cases the sets $\mathcal{M}_{n}
(\mathbf{M}_{n})$ from \eqref{Theor1} and $\mathcal{V}_{n}^{(0)}
(\mathbf{M}_{n})$ from \eqref{defV0} may coincide. Indeed, assume
for model \eqref{mod1} that $\mathbf{M}_{n}=c\mathbf{I}_{n}$,
where $c>0$. Then for any $\mathbf{V}_{n},$  matrices $\mathbf{M}_{n}$ and
$\mathbf{V}_{n}$ commutate, i.e.
$\mathbf{M}_{n}\mathbf{V}_{n}=\mathbf{V}_{n}\mathbf{M}_{n}$.
In this case, Theorems 1 and 2 give
\begin{equation}\label{example1d}
\begin{gathered}
{\cal M}_{n}(\mathbf{M}_{n}) = {\cal V}_{n}^{(0)}(\mathbf{M}_{n}).
\end{gathered}
\end{equation}

{\bf Example 2}. For model \eqref{mod2}, assume that the signal
process $\{s_{i}\}$ has a time-invariant structure
\begin{equation}\label{example2}
\begin{gathered}
s_{i+1} =as_{i} + \sqrt{1-a^{2}}\ u_{i},  \quad i=1,\ldots,n, \\
s_{1} \sim {\cal N}(0,1), \qquad u_{i} \sim {\cal N}(0,1),
\end{gathered}
\end{equation}
where $a$ is a  known scalar (innovation rate parameter) such that
$0\leq a<1$. We assume that the process noise $\{u_{i}\}$ is
independent of the measurement noise $\{\xi_{i}\}$ and the initial
state $s_{1}$ is independent of $\{u_{i}\}$ for all $i$. The
signal sequence $\{s_{i}\}$ corresponds to the auto-regression
process AR(1) and forms a stationary process. This is the main
example considered in \cite{ZhangPoor11, SungTongPoor06}.
The spectra of the observation process under $\mathcal{H}_{0}$ and
$\mathcal{H}_{1}$ are given by
\begin{equation}\label{example2a}
\begin{gathered}
f^{(0)}(\omega) =1,
\ f^{(1)}(\omega) = 1 + f_{\mathbf{S}_{n}}(\omega), \quad
\omega \in [-\pi,\pi],
\end{gathered}
\end{equation}
where the signal spectrum is given by the Poisson kernel
\begin{equation}\label{example2b}
\begin{gathered}
f_{\mathbf{S}_{n}}(\omega) =\frac{1-a^{2}}
{1-2a\cos \omega + a^{2}}.
\end{gathered}
\end{equation}
Then ${\cal F}_{n}^{(1)}(\mathbf{S}_{n})$ takes the form \eqref{stat1},
where (as $n \to \infty$)
\begin{equation}\label{ZP1a}
\begin{gathered}
\int\limits_{-\pi}^{\pi}\ln\left[1+
\frac{f_{\mathbf{S}_{n}}(\omega)
[f_{\mathbf{K}_{n}}(\omega)-f_{\mathbf{S}_{n}}(\omega)]}
{(1+f_{\mathbf{S}_{n}}(\omega))^{2}}\right]d\omega \geq o(1).
\end{gathered}
\end{equation}



\begin{center}{\large  REFERENCES} \end{center}

\begin{enumerate}
\bibitem{Wald}
A.~Wald, \emph{Statistical Decision Functions}. New York: Wiley,
1950.
\bibitem{Lehmann}
E.~L. Lehmann, \emph{Testing of Statistical Hypotheses}.
New York: Wiley, 1959.
\bibitem{Kullback}
S.~Kullback, \emph{Information Theory and Statistics}. New York:
Wiley, 1958.
\bibitem{Stein}
C.~Stein, ``Information and comparison of experiments,'' \emph{unpublished.}
\bibitem{Chernov56}
H.~Chernoff, ``Large-sample theory: parametrics case,''
\emph{Ann. Math. Statist.}, vol. 27, pp. 1-22, 1956.
\bibitem{Poor}
H.~V. Poor, \emph{An Introduction to Signal Detection and
Estimation}, 2nd~ed. New York: Springer-Verlag, 1994.
\bibitem{ZhangPoor11}
W.~Zhang and H.~V. Poor, ``On Minimax Robust Detection of Stationary
Gaussian Signals in White Gaussian Noise,''
\emph{IEEE Trans. Inf. Theory}, vol. 57, no. 6, pp. 3915-3924,
June 2011.
\bibitem{CT}
T.~M. Cover and J.~A. Thomas, \emph{Elements of Information Theory}.
New York: Wiley, 1991.
\bibitem{Bellman}
R.~Bellman, \emph{Introduction to Matrix Analysis}. New York:
McGraw-Hill, 1960.
\bibitem{Horn}
R.~A. Horn and C.~R. Johnson, \emph{Matrix Analysis}. Cambridge:
University Press, 1985.
\bibitem{GS}
U.~Grenander and G.~Szeg\"{o}, \emph{Toeplitz Forms and Their
Applications}. Berkeley: University of California Press, 1958.
\bibitem{SungTongPoor06}
Y.~Sung, L.~Tong and H.~V. Poor, ``Neyman--Pearson Detection of
Gauss--Markov Signals in Noise: Closed-Form Error Exponent and
Properties,'' \emph{IEEE Trans. Inf. Theory}, vol. 52, no. 4,
pp. 1354-1365, April 2006.
\bibitem{Bur17a}
M.~V. Burnashev, "On Detection of Gaussian Stochastic Sequences,"
 \emph{Probl. of Inform. Trans.}, vol. 53, no. 4, pp. 47-66, 2017.
\bibitem{Bur79}
M.~V. Burnashev, ``On the Minimax Detection of an Inaccurately
Known Signal in a White Gaussian Noise Background,''
\emph{Theory of Prob. and Its Appl.}, vol. 24, no. 1, pp. 106-118,
1979.
\bibitem{Bur17}
M.~V. Burnashev, ``Two Comparison Theorems on Distribution of
Gaussian \\
Quadratic Forms,'' \emph{Probl. of Inform. Trans.}, vol.
53, no. 3, pp. 3-15, 2017.
\bibitem{Gray}
R.~M. Gray, ``On the Asymptotic Eigenvalue Distribution of Toeplitz
Matrices,'' \emph{IEEE Trans. Inf. Theory}, vol. 18, no. 6,
pp. 725-730, November 1972.
\bibitem{Petrov}
V.~V. Petrov, \emph{Sums of Independent Random Variables}.
New York: Springer, 1975.
\bibitem{Bur21}
M.~V. Burnashev, ``On Neyman-Pearson Minimax Detection of Poisson
Process Intensity,'' \emph{Statistical Inference for Stochastic
Processes}, vol. 21, no. 1, pp. 211--221, 2021.
\end{enumerate}
\end{document}